\documentclass[preprint,12pt,authoryear]{elsarticle}

\usepackage{amssymb}
\usepackage{url}

\usepackage{hyperref}

\usepackage{etoolbox}
\makeatletter
\patchcmd{\ps@pprintTitle}
  {Preprint submitted}
  {Preprint accepted}
  {}{}
\makeatother

\def\apjl{{ApJ}}                
\def\apjs{{ApJS}}               
\def\ao{{Appl.~Opt.}}           
\def\aap{{A\&A}}                

\usepackage[dvipsnames]{xcolor}

\usepackage[normalem]{ulem}
\usepackage{soul}

\journal{New Astronomy}

\begin{document}

\begin{frontmatter}



\title{SNAD Transient Miner: Finding Missed Transient Events in ZTF DR4 using k-D trees}


\author[inst1,inst2]{P.~D.~Aleo}
\ead{paleo2@illinois.edu}
\affiliation[inst1]{organization={Department of Astronomy, University of Illinois at Urbana-Champaign},
            addressline={1002 West Green Street}, 
            city={Urbana},
            postcode={61801}, 
            state={IL},
            country={USA}}
            
\affiliation[inst2]{organization={Center for AstroPhysical Surveys (CAPS) Fellow, National Center for Supercomputing Applications},
            country={USA}}

\author[inst1,inst3]{K.~L.~Malanchev}

\affiliation[inst3]{organization={Lomonosov Moscow State University, Sternberg Astronomical Institute},
            addressline={Universitetsky pr. 13}, 
            city={Moscow},
            postcode={119234}, 
            country={Russia}}
            
\author[inst3]{M.~V.~Pruzhinskaya}

\author[inst4]{E.~E.~O.~Ishida}

\affiliation[inst4]{organization={Universit\'e Clermont Auvergne, CNRS/IN2P3, LPC},
            city={Clermont-Ferrand},
            postcode={F-63000}, 
            country={France}}
            
\author[inst4]{E.~Russeil}

\author[inst3,inst5]{M.~V.~Kornilov}

\affiliation[inst5]{organization={National Research University Higher School of Economics},
            addressline={21/4 Staraya Basmannaya Ulitsa}, 
            city={Moscow},
            postcode={105066}, 
            country={Russia}}

\author[inst6,inst7]{V.~S.~Korolev}

\affiliation[inst6]{organization={Central Aerohydrodynamic Institute},
            addressline={1 Zhukovsky st}, 
            city={Zhukovsky, Moscow Region},
            postcode={140180}, 
            country={Russia}}
            
\affiliation[inst7]{organization={Moscow Institute of Physics and Technology},
            addressline={9 Institutskiy per.}, 
            city={Dolgoprudny, Moscow Region},
            postcode={141701}, 
            country={Russia}}

\author[inst8]{S.~Sreejith}

\affiliation[inst8]{organization={Physics Department, Brookhaven National Laboratory},
            city={Upton, NY},
            postcode={11973}, 
            country={USA}}

\author[inst9]{A.~A.~Volnova}

\affiliation[inst9]{organization={Space Research Institute of the Russian Academy of Sciences (IKI)},
            addressline={84/32 Profsoyuznaya Street}, 
            city={Moscow},
            postcode={117997}, 
            country={Russia}}
            
\author[inst1,inst10]{G.~S.~Narayan}

\affiliation[inst10]{organization={Center for AstroPhysical Surveys (CAPS), National Center for Supercomputing Applications},
            addressline={1205 West Clark Street}, 
            city={Urbana},
            postcode={61801}, 
            state={IL},
            country={USA}}

\begin{abstract}
We report the automatic detection of 11 transients (7 possible supernovae and 4 active galactic nuclei candidates) within the Zwicky Transient Facility fourth data release (ZTF DR4), all of them observed in 2018 and absent from public catalogs. 
Among these, three were not part of the ZTF alert stream. Our transient mining strategy employs 41 physically motivated features extracted from both real light curves and four simulated light curve models (SN~Ia, SN~II, TDE, SLSN-I). These features are input to a k-D tree algorithm, from which we calculate the 15 nearest neighbors. After pre-processing and selection cuts, our dataset contained approximately a million objects among which we visually inspected the 105 closest neighbors from seven of our brightest, most well-sampled simulations, comprising 89 unique ZTF DR4 sources. Our result illustrates the potential of coherently incorporating domain knowledge and automatic learning algorithms, which is one of the guiding principles directing the \texttt{SNAD} team. It also demonstrates that the ZTF DR is a suitable testing ground for data mining algorithms aiming to prepare for the next generation of astronomical data.
\end{abstract}




\end{frontmatter}


\section{Introduction}
\label{sec:sample1}

The volume and complexity of astronomical data have  drastically increased with the arrival of large scale astronomical surveys, the state of the art being the Zwicky Transient Facility\footnote{\url{https://www.ztf.caltech.edu/}} (ZTF), which generates $\sim$1.4 TB of data per night of observation \citep{Graham2019}. This new data paradigm has forced astronomers to search for automatic tools which can enable classification and discovery within such large datasets.

Traditional supervised machine learning algorithms rely on the availability of large representative training samples, which are not feasible in astronomy \citep{Ishida2019Nat}. Classification requires spectroscopic confirmation, which is an expensive and time consuming process. Given the advent of large scale photometric astronomical surveys, the community has devoted significant resources to systematic spectroscopic follow up of transient candidates. Initiatives like the Young Supernova Experiment \citep[YSE, ][]{Jones2021}, the Bright Transient Survey \citep[BTS, ][]{Fremling2020} and the Global Supernova Project \citep{howell2019} aim to spectroscopically confirm a large number of candidates, which can latter be used as training sets for machine learning applications. Such efforts resulted in a significant increase in the number of available classifications in the last few years. Despite this important advancement, spectroscopic confirmation continues to be a rare commodity due to the impossibility to follow-up all discovered photometric transients. Moreover, observations conditions  required by spectroscopy limit the completeness of such samples for fainter sources, leaving a large fraction of photometrically observed candidates without proper classification. 

In this context, the use of machine learning algorithms opens the possibility of studying larger populations within each class as well as their application in further scientific analysis. A particularly well-studied example of this scenario is the task of supernova (SN) photometric classification. Since their first use as standardizable candles \citep{Riess1998,Perlmutter1999} a lot of effort has been devoted to the development, and adaptation, of machine learning classifiers which may enable purely photometric supernova cosmology \citep[see][and references therein]{Ishida2019Nat}. This covers a large variety of learning algorithms \citep[e.g.][]{Lochner2016,Boone2019,Sooknunan2021, Alves2021}, including deep \citep[e.g.,][]{Muthukrishna2019,Pasquet2019,Moller2020,Villar2020, Allam2021, Burhanudin2021} and adaptive \citep[e.g.,][]{Ishida2019b, Kennamer2020} learning techniques. These deep learning approaches leverage recurrent neural networks \citep[RNNs,][]{Muthukrishna2019, Moller2020}, RNN-based autoencoders \citep{Sadeh2020, Villar2021}, Temporal Convolutional Networks \citep[TCNs,][]{Muthukrishna2021}, and more.

This new direction towards data driven approaches has also benefited from increasingly more realistic simulations. Given the sparse availability of confirmed classifications, simulations have filled the gap when real data are not available. They have been used to compare results from different classifiers in the context of data challenges, like SNPhotCC \citep{Kessler2010} and PLAsTiCC \citep{Hlozek2020}, as well as used to replace training samples in transfer learning scenarios \citep[e.g.,][]{Pasquet2019}. In all such attempts, large volumes of simulations are used to infer statistical properties of different classes which can allow bulk data classification. In parallel, unsupervised learning techniques have been used for clustering \citep[e.g.,][]{KroneMartins2014, Ralph2019,Pera2021} and anomaly detection \citep[e.g.,][]{Villar2020, Fisher2021, galarza2021, lochner2021}
without the need of labels. In this context, simulations have also been used to validate the proposed algorithms \citep[e.g., ][]{Villar2021}. 
 
The \texttt{SNAD}\footnote{\url{https://snad.space/}} team has been continuously working in the development of anomaly detection algorithms which are able to prove their efficiency in real data while incorporating domain knowledge in the machine learning model -- thus tailoring it according to the scientific interest of the expert \citep[e.g.,][]{Pruzhinskaya2019,Aleo2020,Malanchev2021, ishida2021}. In this work, we present a hybrid approach for mining transients in large astronomical datasets, specifically ZTF DR4; moreover, our methodology can also be applied to the nightly ZTF alert-stream via time-domain brokers like ANTARES \citep{Matheson2021} and FINK \citep{Moller2021}. Ultimately, our analysis still focuses on performance on real data; however, we use a few simulated light curves as a guide to identify transients in a large dataset mostly comprised of variable stars~\citep{2020ApJS..249...18C}.

We describe the data, simulations, and pre-processing steps in Section \ref{sec:data}. The simulation-to-data matching algorithm is described in Section \ref{sec:method} and results are shown in Section \ref{sec:res}. Our conclusions are outlined in Section \ref{sec:conclusions}.

\section{Data}
\label{sec:data}

\subsection{ZTF DR4}
\label{subsec:ztf}

ZTF is a northern sky survey stationed at Mount Palomar. It  uses a 48-inch Schmidt telescope equipped with a 47~deg$^{2}$ camera, with primary science objectives including the physics of SN and relativistic explosions, multi-messenger astrophysics, SN cosmology, active galactic nuclei (AGN), tidal disruption events (TDE), stellar variability and Solar System objects \citep{Graham2019}. The survey started on March 2018 and, during its initial phase,  has observed around a billion objects \citep{Bellm2019} in three photometric bands: ZTF-$g$, ZTF-$r$, ZTF-$i$. It is employed as a testing ground for the next generation of large scale surveys like the Vera Rubin Observatory Legacy Survey of Space and Time (LSST,~\citealt{2009arXiv0912.0201L}). For this work, we use the survey's fourth data release (DR4), made public on December 9, 2020\footnote{\url{https://web.ipac.caltech.edu/staff/fmasci/ztf/dr4/ZTF_DR4.pdf}}. In this work we use the two bluer filters, ZTF-$r$ and ZTF-$g$, and denote them as $zr-$ and $zg-$.

\subsection{Simulations}
\label{subsec:sims}

We create realistic ZTF simulations using SNANA \citep{Kessler2009}, a catalog-level light curve simulator which includes effects due to telescope characteristics and observational conditions. 
We adopt ZTF data release~3 cadence and magnitude error distribution 
(see \citealt{Chatterjee2021} for details). Starting from the template models originally developed for the Photometric LSST Astronomical Time-series Classification Challenge \citep[PLAsTiCC, ][]{Kessler2019, Hlozek2020}, we selected those whose light curve were clearly non-periodic and/or with largest probability of being detected (high intrinsic brightness). Thus, our final model sample contained only SN~Ia, SN~II, SLSN-I, and TDE. 
\par

\begin{figure}[h]
\centering
\includegraphics[width=0.65\textwidth, scale=1]{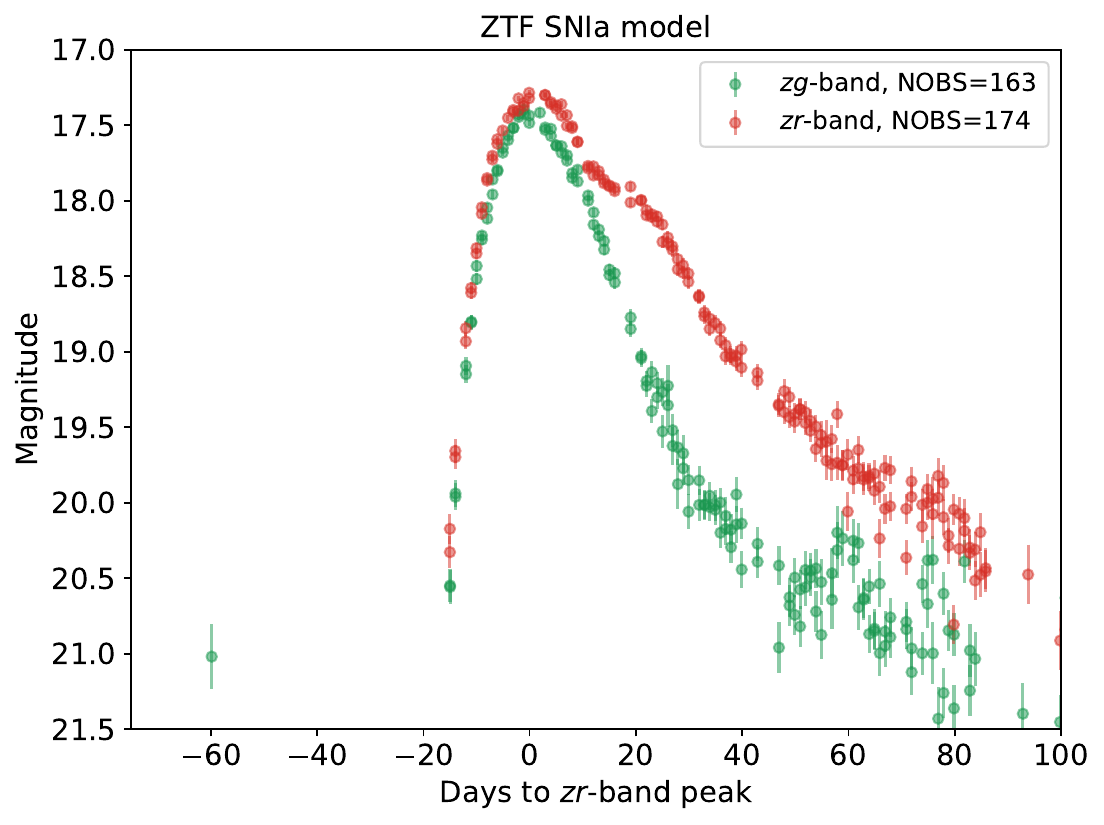}
\caption{SNANA simulation of an SN Ia used to match with ZTF DR4 data. Green and red circles correspond to $zg-$ and $zr-$band observations, respectively. Both bands have more than 100 detections with SNR $\textgreater$ 5. This simulation's tenth nearest neighbor via the k-D tree algorithm is {\tt SNAD156}/AT~2018lzb.}
\label{fig:snia_sim}
\end{figure}

For each model, we generated $\sim$50,000\footnote{This number results in a small but sufficient portion of well-sampled simulations located in the bright tail of the peak magnitude distribution.} simulations and imposed a series of quality cuts (SNR $\textgreater$ 5, magnitude limit, $m < 21.5^m$, and a minimum of 100 observations in each of the $zr-$ and $zg-$bands). We then chose, among the surviving objects, the seven brightest light curves (3 SLSN-I, 1 SN Ia, 1 SN II and 2 TDE with peak magnitude $\sim$17$^m$) and used these objects as input to our k-D tree (see Section~\ref{sec:method}). 
An example of a ZTF SN Ia simulation is given in Fig.~\ref{fig:snia_sim}. \par


\subsection{Light curve selection}
\label{subsec:xmatch}

ZTF DR4 includes a photometric dataset grouped into objects by a detection position, an observation field, and a passband. Thus, a single stellar source can be represented by multiple objects due to overlapping fields and several observed passbands.
We perform $0.2''$ cross-matching to associate different objects as a single source using the ClickHouse database management system\footnote{\url{https://clickhouse.com}}.
We select observations with \texttt{catflags $=0$}, positive \texttt{magerr}, in $zg-$ \& $zr-$passbands only, and within the first 420 days of the survey. Moreover, we consider objects with the absolute value of the galactic latitude larger than 15 degrees and having at least 100 detections per passband (non-detections are not presented in ZTF~DRs).
Finally, we filter out light curves having  small variability for which  $\frac1{N-1} \sum (\frac{m_i - \bar{m}}{\sigma_i})^2 < 3$, where $N$ is the number of observations, $m_i$ and $\sigma_i$ are detection magnitudes and corresponding error estimates, and $\bar{m} \equiv \sum(m_i/\sigma_i^2) / \sum(1/\sigma_i^2)$ is a weighted mean magnitude. As a result, we were left with $990,220$ sources.  

\subsection{Pre-processing}
\label{subsec:preproc}

We extract light curve features for both simulations and our sample of ZTF DR4 data with the \texttt{light-curve}\footnote{\url{https://github.com/light-curve}} package \citep{Malanchev2021}. We use a total of 82 light curve features without scaling or normalization (41 per band) including magnitude amplitude, Stetson $K$~coefficient \citep{Stetson1996}, standard deviation of Lomb-Scargle periodogram \citep{1976Ap&SS..39..447L,1982ApJ...263..835S}, etc. Our list encloses 68 magnitude-based and 14 flux-based features, whereby 66 are brightness-related and 16 are temporally-related; see the complete list in \ref{sec:features}.


\section{Methodology}
\label{sec:method}

We take seven of the brightest simulations (3 SLSN-I, 1 SN Ia, 1 SN II and 2 TDE with peak magnitude $\sim$17$^m$) and apply a k-D tree \citep{Bentley1975} to their extracted 82 non-normalized features.\footnote{Generally, feature scaling can be used to ensure all single-dimensional distances map to the same range of values. This is done in order to equally weight the calculation of k-D distances across all features. In our case, the chosen feature set already meets this condition. In trying to apply the same method for another feature set, the reader should be aware of this condition.} Subsequently, we identify the 15 nearest neighbors for each simulation (105 matches in total, resulting in 89 unique ZTF DR4 sources). We chose to only inspect the first 15 nearest neighbors because it enables a sufficient test of the method's validity while providing a manageable number of light curves to visually examine.

We inspected their light curves using the \texttt{SNAD} ZTF viewer\footnote{\url{https://ztf.snad.space/}} \citep{Malanchev2021}, a tool that allows easy access to the individual exposure images; to the Aladin Sky Atlas~\citep{2000A&AS..143...33B,2014ASPC..485..277B}; and to various catalogues of variable stars and transients, including the General Catalogue of Variable Stars (GCVS,~\citealt{2017ARep...61...80S}), the American Association of Variable Star Observers' Variable Star Index (AAVSO VSX,~\citealt{2006SASS...25...47W}), the Asteroid Terrestrial-impact Last Alert System (ATLAS,~\citealt{2018AJ....156..241H}), the ZTF Catalog of Periodic Variable Stars~\citep{2020ApJS..249...18C}, astrocats\footnote{\url{https://astrocats.space/}}, the OGLE-III On-line Catalog of Variable Stars~\citep{Soszynski2008}, and the  \textsc{SIMBAD} database~\citep{2000A&AS..143....9W}. It also contains the information about object colour and galactic extinction in the chosen direction. 
Thus, we consider that all publicly available sources were included in our search. 
If an object was not announced by a survey, we consider it as undiscovered.


The k-D tree takes $\sim$4 seconds on a 2 GHz Quad-Core Intel Core i5 processor. Although we use the default Euclidean distance metric for distance, we will explore other distance metrics such as Manhattan and Mahalanobis distances in future work.

The large dimension of the resulting parameter space is prone to produce non-optimal efficiency when presented to the k-D tree algorithm (curse of dimensionality). This is especially important when there is some correlation among the features \citep[see][Figure A.3]{Malanchev2021}. In order to access the performance of our method in a lower dimension parameter space, we also perform a similar analysis using a principal component analysis \citep[PCA,][]{jolliffe2013principal} approach and compare results with those obtained from the complete feature set (see Section \ref{sec:conclusions}). 




\begin{table}
\caption{Results of transient mining in ZTF DR4 using k-D trees.}
\begin{scriptsize}
\begin{tabular}{lccccccl}
\hline
\hline
Name	&	R.A.	&	Dec. & Host galaxy$^*$ & $z_{\rm ph}$	&	TNS	&	Type$^\dagger$	&	Comments	\\
\hline
{\tt SNAD149}	&	284.35219	&	67.17743 & &	&	AT 2018lyu	&	PSN	& ZTF18abrlhnm		\\
{\tt SNAD150}	&	240.75660	&	32.64360 & SDSS J160301.52+323835.8 &	&	AT 2018lyv	&	PSN	&	ZTF18aaqnsia	\\
{\tt SNAD151}	&	190.61613	&	52.77285 & SDSS J124227.87+524621.9 & 0.20$\pm$0.11	&		AT 2018lyw	&	PSN	&	ZTF19aaoykyz	\\
{\tt SNAD152}	&	212.40042	&	55.95669 & SDSS J140936.12+555724.5 & 0.11$\pm$0.03	&	AT 2018lyx	&	PSN	&	ZTF18aanaryv	\\
{\tt SNAD153}	&	325.96473	&	24.35382 & SDSS J214351.53+242113.6 & 0.17$\pm$0.04	&	AT 2018lyy	&	AGN	&	ZTF18abvosry	\\
{\tt SNAD154}	&	219.40247	&	38.04782 & SDSS J143736.57+380252.0 & 0.22$\pm$0.08	&		AT 2018lyz	&	AGN	&		\\
{\tt SNAD155}	&	212.29492	&	38.70615 & SDSS J140910.77+384222.1 & 0.39$\pm$0.06	&		AT 2018lza	&	AGN	&	ZTF18acwyyib	\\
{\tt SNAD156}	&	184.83691	&	45.49015 & SDSS J121920.85+452924.4 & 0.09$\pm$0.02	&		AT 2018lzb	&	PSN	&		\\
{\tt SNAD157}	&	280.69720	&	36.36783 & &		&	AT 2018lzc &	PSN	&	ZTF18abegwmh	\\
{\tt SNAD158}	&	253.15763	&	25.82260 & SDSS J165237.84+254921.2 & 0.14$\pm$0.05	&		AT 2018lzd	&	AGN	&		\\
{\tt SNAD159}	&	216.86359	&	53.08862 & SDSS J142727.25+530518.9 & 0.17$\pm$0.07	&		AT 2018lze	&	PSN	&	ZTF19aanjzps	\\
\hline
\end{tabular}
\label{tab:tr_cand}\\
\begin{flushleft}
$^*$
If available, candidate host galaxies from SDSS DR16~\citep{2020ApJS..249....3A} and their corresponding photometric redshifts ($z_{\rm ph}$).\\
$^\dagger$
PSN --- possible supernova, AGN --- active galactic nucleus.
\end{flushleft}
\end{scriptsize}
\end{table}

\section{Results}
\label{sec:res}

\subsection{Supernova and AGN candidates}

As a result of transient mining, we discovered 11 previously unreported supernova and active galactic nucleus candidates (see Table~\ref{tab:tr_cand}). The remaining 94 matches (81 unique ZTF DR4 sources) were either known/already reported transients or variable stars. The full breakdown of all uniquely-matched source objects is presented in Table~\ref{tab:match_obj}. We note that
$\sim 50\%$ of our matched sources are variable stars (44), and 18 were transients. 
Given the large number of variable stars previously estimated in ZTF data releases \citep{2020ApJS..249...18C}, and our imposed selection cuts 
(see Section~\ref{subsec:xmatch}), it is reasonable to expect  
that a large fraction of our data set is composed of 
well sampled, high-amplitude variable stars whose coverage in parameter space significantly overlaps with the regions populated by transients  
(see, e.g. Figure~\ref{fig:magn_amp_scatter}). 
Thus, a ratio of 18 transients (11 newly-discovered) to 44 variable stars out of 89 unique sources selected from 990,220 considered ZTF sources is a very successful result. Considering the extreme case where $\approx 3000$ SNe discovered by ZTF \citep{dhawan2022} were part of our data set, the expected incidence of SNe when choosing 100 sources at random would be of $< 1 (\approx 0.3)$ event.


Table~\ref{tab:transient_demo} shows a further breakdown of these 18 transients, detailing the 7 previously known and 11 \texttt{SNAD}-discovered transients from this paper, as well as  
the $k$-nearest neighbours match to its simulation type. Although we cannot claim that every instance of model neighbor will define the class of the found transient (particularly when there are several model simulation types matched to a singular event, e.g. AT\,2019gsn, \texttt{SNAD149}, \texttt{SNAD150}, \texttt{SNAD151}), we note that in the two cases where the spectroscopic class was available, it did match the simulation class: SN\,2018aej, SN\,2018aoy. Any possible correlation 
will be explored in future work.

\begin{table}
\centering
\caption{Total uniquely-matched ZTF DR4 sources, first 15 $k$-nearest neighbors.}
\begin{scriptsize}
\begin{tabular}{c|c}
\hline
\hline
Matched Source Type (count)  &	Subtype(s) (count)/Other notes	\\
\hline
Variable Star (44)  &  RR Lyrae (27), UG (10), Mira (3), Semi-Regular (2), Cepheid (1), RS CVn (1)  \\
Quasi-stellar (15)	&	QSO (14),  BL Lacertae (1)  \\
Galaxy (3)	&	Seyfert (2),  Radio source (1) 	\\
Transient$^\dagger$ (18) &  Reported w/ classification (2), Reported w/o classification (5), \textbf{This work (11)} \\
Uncatalogued (9)    &	\\
Artifact (0)	&		\\
\hline
\end{tabular}
\label{tab:match_obj}\\
\begin{flushleft}
$^\dagger$ A breakdown of this category, including various demographics with nearest-neighbor and simulation type matches, can be found in Table~\ref{tab:transient_demo}.
\end{flushleft}
\end{scriptsize}
\end{table}
\begin{table*}
\caption{Transient demographics among the first 15 $k$-nearest neighbors.}
\begin{scriptsize}
\begin{tabular}{ccccc}
\hline
\hline
Name/TNS    &   $k$-nearest neighbor	&	Previously Known?   &   Matched Simulation    &   Class (if known) \\
\hline
ATLAS20hzm/AT 2019mzp	&   $k$=2    &     Yes    &	SLSN-I  &    \\
ATLAS19mbr/AT 2019gsn	&   $k$=3    &     Yes    &	SN~II   &    \\
	''  &   $k$=5    &        &    TDE  &    \\
	''  &   $k$=8    &         &    SN~Ia    &    \\
	''  &   $k$=8    &         &    SLSN-I   &    \\
SNAD129/AT 2018lxn  &   $k$=9    &     Yes    &    SN~Ia    &    \\
ATLAS18oay/AT 2018bhq  &   $k$=12    &     Yes    &    TDE    &    \\
PS18mh/SN 2018aej  &   $k$=13    &     Yes    &    SN~Ia    &  SN~Ia  \\
ZTF19aaqejgh/AT 2018dvs  &   $k$=14    &     Yes    &    SN~Ia    &   \\
ATLAS18mzo/SN 2018aoy   &   $k$=14    &     Yes    &    SN~Ia    &   SN~Ia \\
\hline
{\tt SNAD149}/AT 2018lyu	&     $k$=2    &    	No	&	TDE &    \\
        ''     &     $k$=12    &    	&	SLSN-I &    \\
{\tt SNAD150}/AT 2018lyv	&     $k$=4    &    	No	&	TDE &    \\
        ''      &     $k$=5    &    	&	SN~Ia &    \\
        ''      &     $k$=6    &    	&	SLSN-I &    \\
{\tt SNAD151}/AT 2018lyw	&     $k$=6    &    	No	&	TDE &    \\
        ''     &     $k$=8    &    	&	SN~Ia &    \\
{\tt SNAD152}/AT 2018lyx	&     $k$=2    &    	No	&	SN~Ia &    \\
{\tt SNAD153}/AT 2018lyy	&     $k$=6    &    	No	&	TDE &    \\
{\tt SNAD154}/AT 2018lyz	&     $k$=13    &    	No	&	TDE &    \\
{\tt SNAD155}/AT 2018lza	&     $k$=9    &    	No	&	TDE &    \\
{\tt SNAD156}/AT 2018lzb	&     $k$=10    &    	No	&	SN~Ia &    \\
{\tt SNAD157}/AT 2018lzc	&     $k$=11   &    	No	&	SN~Ia &    \\
{\tt SNAD158}/AT 2018lzd	&     $k$=14    &    	No	&	SN~II &    \\
{\tt SNAD159}/AT 2018lze	&     $k$=15   &    	No	&	SN~II &    \\
\hline
\hline
\end{tabular}
\label{tab:transient_demo}\\
\begin{flushleft}
Note: Only unique matches have their official name listed. Additional (duplicate) matches for any particular source is subsequently denoted by quotation marks. 
\end{flushleft}
\end{scriptsize}
\end{table*}

Among the 11 new objects, 9 are coincident with galaxy positions cataloged in SDSS DR16~\citep{2020ApJS..249....3A}, thus confirming their extragalactic origin. We conclude that 7 of the candidates are likely to be SNe and 4 others are AGN candidates (see~Fig.~\ref{fig:snad_LC2}). All candidates were sent to the Transient Name Server\footnote{\url{https://www.wis-tns.org/}} (TNS) and received an official TNS identifier as well as an internal \texttt{SNAD} name (Table~\ref{tab:tr_cand}).

Among our discovery objects, {\tt SNAD150}, {\tt SNAD152}, and {\tt SNAD155} occurred when the ZTF template reference image was taken. This detail provides a possible reason for why they were initially unreported, as the ZTF alert stream utilizes differential photometry, and typically the science image contains the transient, whereas here the reference image contains the transient.
Perhaps more importantly, another three objects from our list are missing in the official ZTF alert stream, two of which we classify as AGN candidates ({\tt SNAD154}, {\tt SNAD158}) and the third as a possible SN ({\tt SNAD156}). Using the \texttt{SNAD} viewer, we investigated the placement of all newly discovered transients in the original FITS files, in order to confirm if this could be a result of missing reference image in chip-edges. We found that {\tt SNAD155} and {\tt SNAD159} are indeed placed somewhat close to image borders, but those are not the objects absent in the alert stream. Missed transients {\tt SNAD154} and {\tt SNAD156} have a peak magnitude $\sim18.5^m-19.5^m$ which is comparable to those of other SNAD objects listed here but are also detected by the alert system. Moreover, some of our candidates (e.g., {\tt SNAD150}, {\tt SNAD151}) have well-sampled early light curves which could help constrain the progenitor parameters of supernovae and shed light on the explosion scenarios \citep{Jones2021}.

To illustrate the scientific significance of these  candidates, we perform light curve fits on two of them, {\tt SNAD150} and {\tt SNAD154}. We use the \textsc{Python} library \textsc{sncosmo}\footnote{\url{https://sncosmo.readthedocs.io/en/stable/}} to fit their light curves with Peter Nugent's spectral templates\footnote{\url{https://c3.lbl.gov/nugent/nugent_templates.html}} which cover the main supernova types (Ia, Ib/c, IIP, IIL, IIn). Nugent's models are simple spectral time series that can be scaled up and down. The zero phase is defined relative to the explosion moment and the observed time $t$ is related to phase via $t = t_0 + {\rm phase} \times (1 + z)$. Other model parameters are redshift $z$, observer-frame time corresponding to the zero source's phase $t_0$, and the amplitude.
 
\begin{figure}[h]
\centering
\includegraphics[width=1\textwidth, scale=1]{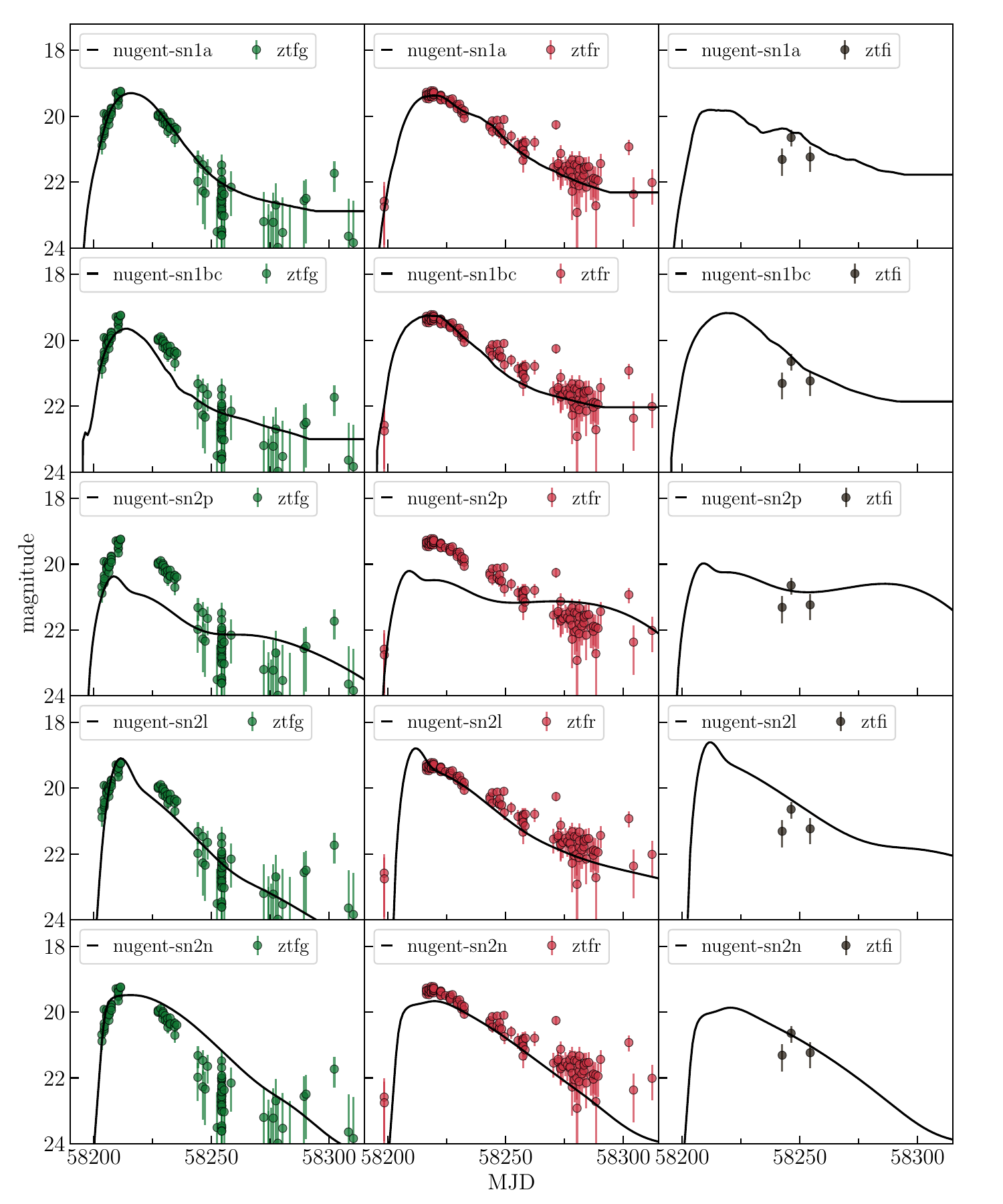}
\caption{Results of light curve fit of {\tt SNAD150} by Nugent's supernova models. Observational data correspond to OIDs: {\tt 679108100003227} ($zg$), {\tt 679208100014706} ($zr$), {\tt 679308100021192} ($zi$).}
\label{fig:snad150}
\end{figure}

\begin{figure}[h]
\centering
\includegraphics[width=1\textwidth, scale=1]{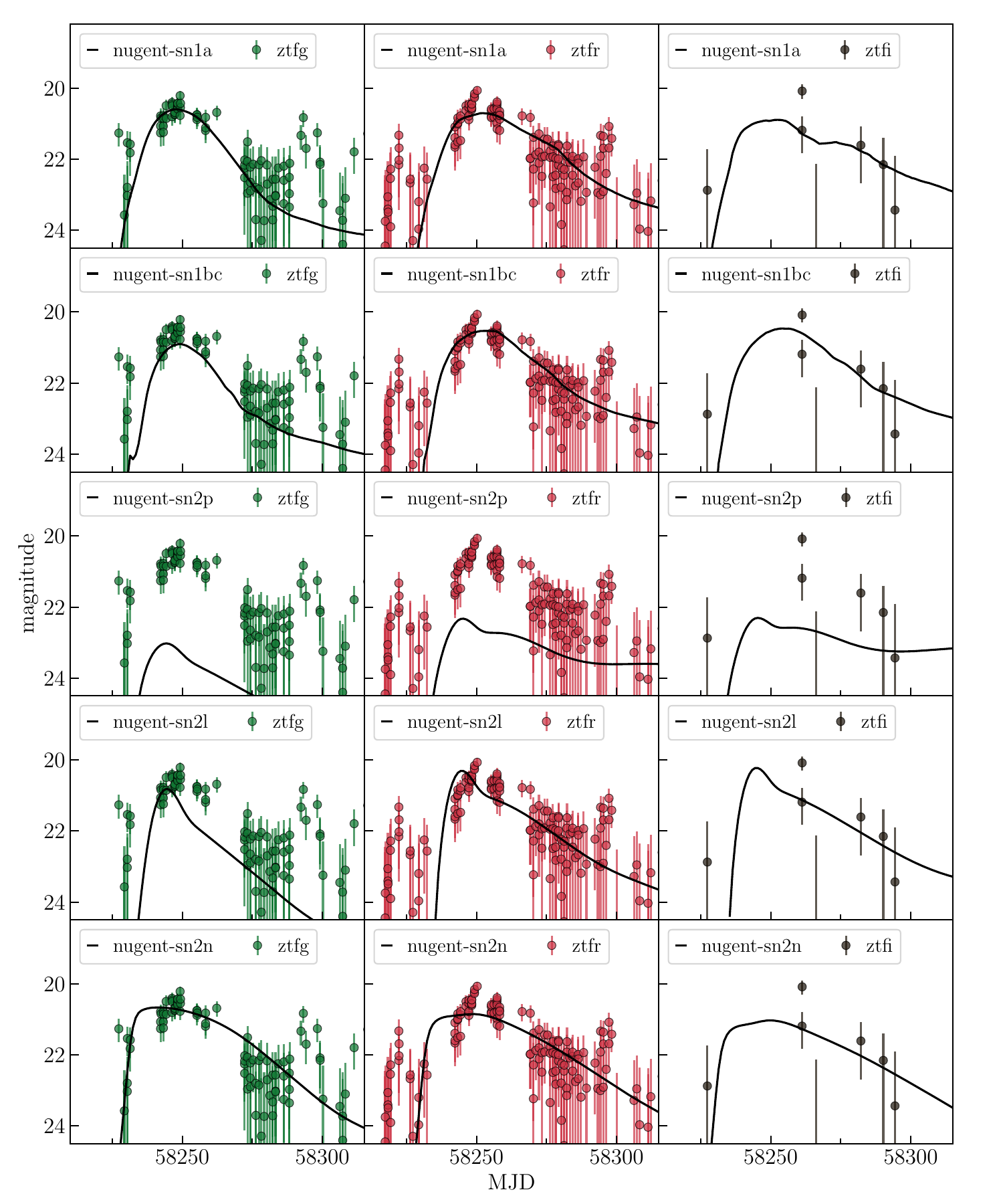}
\caption{Results of light curve fit of {\tt SNAD154} by Nugent's supernova models. Observational data correspond to OIDs: {\tt 719102100006086} ($zg$), {\tt 719202100004008} ($zr$), {\tt 719302100018848} ($zi$).}
\label{fig:snad154}
\end{figure}

We extract photometry in $zg-, zr-$, and $zi-$ passbands from only one field. Then, we subtract the reference magnitude from ZTF light curves to roughly account for the host galaxy contamination. To estimate the redshift bounds for {\tt SNAD150}, we adopt $[-15; -22]$ as an acceptable region for the supernovae absolute magnitude \citep{2014AJ....147..118R} and then, using the apparent maximum magnitude, transform it to the possible redshift range. For {\tt SNAD154} there is a known SDSS galaxy at the source position with measured photometric redshift and corresponding errors, which we use for redshift bounds. Results of the fit are shown in Figs.~\ref{fig:snad150} and \ref{fig:snad154}. {\tt SNAD150} light curves are best described by Nugent's Type Ia Supernova model, while {\tt SNAD154} is not well fitted by any of them---the observed light curve width near the peak brightness in $zg$ and $zr$ bands is smaller than the models suggest. Thus, we conclude that {\tt SNAD154} object is likely to be an active galactic nucleus.



\subsection{Feature space examination}
\label{subsec:feat_space}

We also investigate the effectiveness of simple data-quality cuts in discovering the {\tt SNAD} transients listed in this work, when compared to the results obtained from the complete pipeline using an heuristic search. 
Using the assumption that such transients could be found by selecting the highest-amplitude objects among a relatively bright population (peak magnitude $\sim18^m$), we selected all objects fulfilling three criteria in both $zg-$ and $zr-$ light curves: 

\begin{enumerate}
    \item \textit{error-weighted mean magnitude} $\in (18^m; 21^m)$, which removes bright and faint light curves with few bright outlier observations,
    \item the highest \textit{Lomb--Scargle periodogram peak power} to the \textit{power standard deviation ratio} $< 10$, which removes light curves showing periodicity,\footnote{Some of the light curves could show bogus variability, e.g. $\sim 1$\,day, so this criterion could remove some non-periodic sources.}
    \item light curve \textit{magnitude amplitude} $>1.75^m$.\footnote{We chose this value to obtain a new set of 105 objects to visually inspect and compare against our 105 matched nearest-neighbors.}
\end{enumerate}
 
After visually inspecting 105 objects resulting from this heuristic search, we found no supernova-like or AGN-like light curves. In fact, most of the objects present artifacts or are standard eruptive and cataclysmic variables.

\begin{figure}[h]
\centering
\includegraphics[width=0.75\textwidth, scale=1]{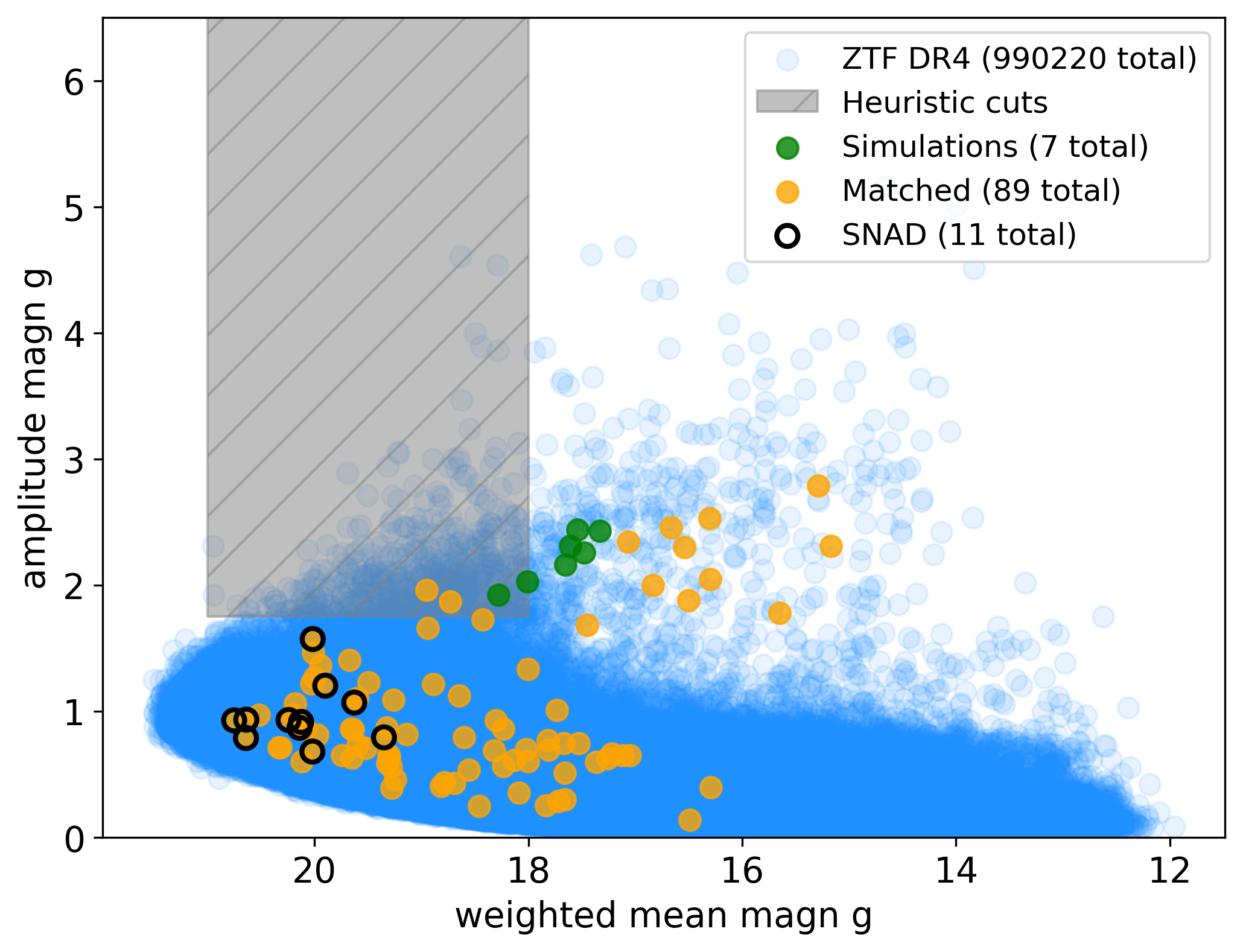}
\caption{A scatterplot of two light curve features (\textit{$zg-$ error-weighted mean magnitude} and \textit{$zg-$ magnitude amplitude}) of all ZTF~DR4 sources considered in this work (blue circles), simulations used as input to the k-D tree (green circles), uniquely matched ZTF~DR4 sources (yellow circles), and those unique sources which constitute our new \texttt{SNAD} transients (outer black circles). We show a hatched region of our heuristic cuts from Sec.~\ref{subsec:feat_space}, highlighting that none of our new \texttt{SNAD} transients could be found with simple data quality cuts.}
\label{fig:magn_amp_scatter}
\end{figure}

Figure~\ref{fig:magn_amp_scatter} shows that cuts in magnitude feature space (here we used $zg-$band \textit{magnitude amplitude} and $zg-$band \textit{error-weighted mean magnitude}) are not able to isolate our {\tt SNAD} transients from the rest of the ZTF DR4 sample, including the region of feature space probed by the heuristic search. In fact, even if one \textit{apriori} knew the magnitude mean and amplitude ranges to look for such transients, there would be an overwhelming amount of other variable stars, QSOs, and other non-transient phenomena among the ZTF DR4 objects to inspect. Considering the density of total sources in the \textit{magnitude amplitude} $\in[0.7; 1.7]$, \textit{error-weighted mean magnitude} $\in[19^m; 21^m]$ region, it is unlikely that one would retrieve our  discovered {\tt SNAD} transients without the machine learning part of the pipeline. \par

\section{Conclusions}
\label{sec:conclusions}

The consequences of large and complex astronomical datasets have been extensively discussed in the literature. A large part of such discussions focus on the potential of different  machine learning algorithms, and are backed by their performance on large simulated datasets. This allows researchers to report on statistical properties of classifiers. In this work, we explored the potential of simulations from a different perspective. 

Our analysis relies on two hypotheses: 1) state of the art simulations are a good proxy to real data, and 2) astrophysically inspired features correctly summarize the information necessary to characterize transient light curves. These hypotheses perfectly translate the underlying principles of all \texttt{SNAD} efforts, whose focus is to construct environments that coherently incorporate domain knowledge in learning strategies.  

Both statements were tested by extracting the same 82 features from  a small number of simulations (7 light curves representing 4 different classes) as well as from real data (ZTF DR4, comprising $990,220$ objects). Subsequently, a k-D tree with Euclidean distance was used to search for the 15 objects in real data which were closest match in feature space to each simulation. By visually inspecting 105 total source matches (89 unique), we identified 11 previously unreported transient events, all occurring in 2018. Among these, 7 possible supernovae and 4 AGN candidates, the majority of them with light curves containing several observed epochs before maximum brightness.

We are aware that at 82 dimensions, the k-D tree suffers from the curse of dimensionality, and in general, using a k-D tree of $k$-dimensions to probe $N$ points in a dataset should satisfy $N \gg 2^{k}$; otherwise, the efficiency of the search is no better than an exhaustive one \citep{Kung2001}. To this point, we are not aiming for the most efficient search algorithm, but an approach allowing us to find closest neighbours in the light-curve feature space. Moreover, because we only had 7 simulations to associate, our k-D tree search only takes $\sim$4 seconds to run.
To further investigate how the efficiency of our method is affected by the high number of features, we also performed the k-D tree analysis in a 15 dimension parameter space resulting from Principal Component Analysis \citep[PCA,][]{Jolliffe1986}. This dimensionality reduction retains 86\% of the variance of our original 82 dimensional parameter space. We obtained results in good agreement (Mira stars, cataclysmic variables, confirmed SN, unconfirmed transients, etc.) with those presented here, including some of the same discovery objects listed in this work (e.g. {\tt SNAD150}). A detailed analysis of this lower dimensional parameter space and its quantitative impact on our results is an interesting investigation suited for subsequent work. 


Beyond confirming our initial hypothesis about the importance of simulations and physically inspired features, such results also highlight the importance of further improving observation pipelines and the central role classical learning algorithms can play in this task. Our candidates include 3 objects which were not part of the ZTF alert stream despite being as bright as the others and holding a considerable number of pre-maximum observations. Although currently it is not feasible to precisely determine why such candidates were lost, a few possible explanations include: the completeness of magnitude limited surveys for fainter sources, active transients when the template image was taken (which may explain {\tt SNAD150, SNAD152} and {\tt SNAD155}) or issues in the imaging pipeline. However, it is likely that the main reason behind the high fraction of non-identified transients ($\approx 12\%$, 11 new sources among 89 visually inspected) is the period of the data scrutinized in this work. All our sources were observed in 2018, when the ZTF ecosystem was in the beginning of its operations. It is reasonable to expect that a few transients were lost due to non-optimal pipeline parameters which were improved in subsequent years. We are currently working on applying the \texttt{SNAD} infrastructure to data from latter years to better understand how the fraction of non-reported transients vary with time. This will shed some light into the reasons behind the results presented in this paper and, at the same time, help prevent similar future losses. Given the importance of early observations in astrophysical investigations, this understanding is an important step towards an optimal exploitation of our observational resources. 

Finally, we highlight the potential of ZTF DRs as a fertile ground for testing machine learning pipelines currently being prepared for future large scale surveys. Since we expect transient events to be a small fraction of the total number of objects in the DRs, it provides a perfect environment to stress-test data mining and anomaly detection algorithms, which do not require a large number of labels, before the arrival of LSST. 

The era of big data in astronomy has imposed the necessity of automatic learning algorithms in order to digest large and complex datasets. Nevertheless, we should not underestimate the role of domain experts when adapting machines to work in real scientific data environments. Their input is vital to direct changes in the learning strategy due to the expert's response (e.g. as in active learning algorithms) or in the design of physically motivated features, realistic simulations and physically motivated evaluation criteria. The successful incorporation of this long acquired knowledge to automatic learning frameworks is necessary to ensure we will be able to fully explore the scientific potential of our data.

\section{Acknowledgments}
\label{sec:acknowledgment}

The authors thank Vasilisa Malancheva, Anastasia Malancheva, and Vadim Krushinsky for enlightening discussions. We would also like to thank Rick Kessler for his help with ZTF simulations and SNANA and the anonymous referee for comments that helped improved the clarity of the paper. The authors thank Jelena Banjac and Jo\'e Veber for logistical help and hosting during the SNAD IV workshop, of which this work is a product.

The reported study was funded by RFBR and CNRS according to the research project № 21-52-15024. SNAD receives financial support from CNRS International Emerging Actions under the project \textit{Real-time analysis of astronomical data for the Legacy Survey of Space and Time} during 2021-2022. The authors acknowledge the support by the Interdisciplinary Scientific and Educational School of Moscow University ``Fundamental and Applied Space Research''. P.D.A. is supported by the Center for Astrophysical Surveys at the National Center for Supercomputing Applications (NCSA) as an Illinois Survey Science Graduate Fellow. This research also used resources of the National Energy Research Scientific Computing Center (NERSC), a U.S. Department of Energy Office of Science User Facility located at Lawrence Berkeley National Laboratory, operated under Contract No.~DE-AC02-05CH11231. A.A.V. is supported by RSFC grant 18-12-00378 for anomalies light curves analysis.

This manuscript has been authored by employees of Brookhaven Science Associates, LLC under Contract No.~DE-SC0012704 with the U.S. Department of Energy. The publisher by accepting the manuscript for publication acknowledges that the United States Government retains a non-exclusive, paid-up, irrevocable, world-wide license to publish or reproduce the published form of this manuscript, or allow others to do so, for United States Government purposes.

This research has made use of NASA’s Astrophysics Data System Bibliographic Services and following Python software packages: {\sc NumPy}~\citep{numpy}, {\sc Matplotlib}~\citep{matplotlib}, {\sc SciPy}~\citep{scipy}, {\sc pandas}~\citep{reback2020pandas,mckinney-proc-scipy-2010}, {\sc scikit-learn}~\citep{scikit-learn}, {\sc astropy}~\citep{astropy:2013,astropy:2018}, and {\sc astroquery}~\citep{2019AJ....157...98G}.



\bibliographystyle{elsarticle-harv} 
\bibliography{main}

\appendix

\section{ZTF DR4 light curves}
\label{sec:appendix}
Light curves of 11 newly found transients, generated with the ZTF SNAD viewer. The plots show only data from ZTF DR4 used for feature extraction (Section~\ref{subsec:preproc}).

\begin{figure*}
    \begin{minipage}{0.49\linewidth}
        \centering
        \includegraphics[scale=0.4]{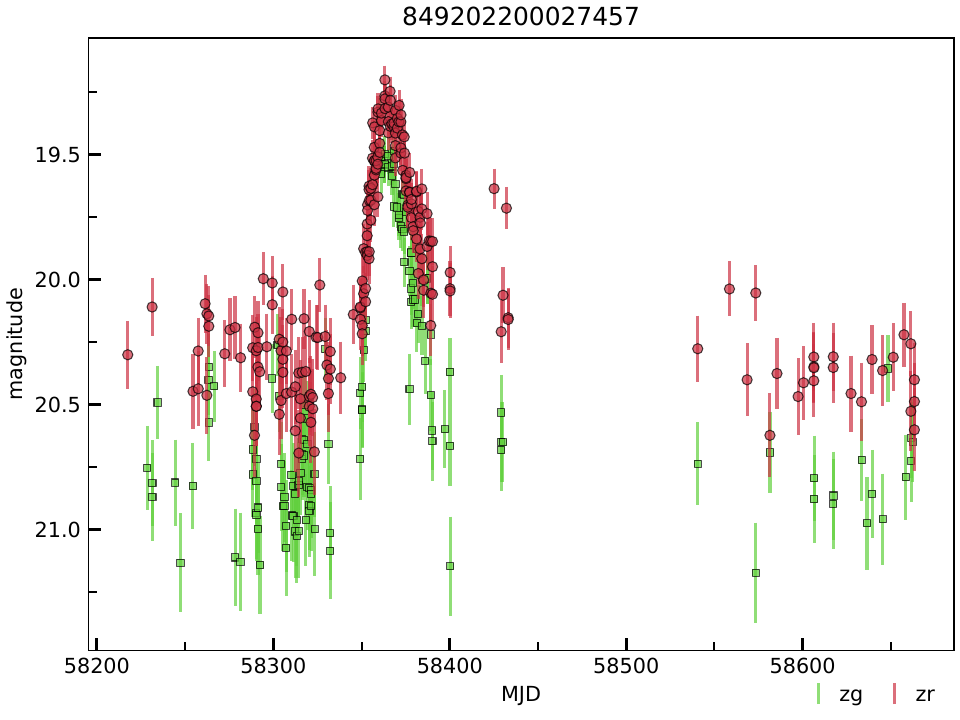}\\a) {\tt SNAD149}: \texttt{849102200015533}~($zg$), \texttt{849202200027457}~($zr$)\\
    \end{minipage}
    \hfill
    \begin{minipage}{0.49\linewidth}
        \centering
        \includegraphics[scale=0.4]{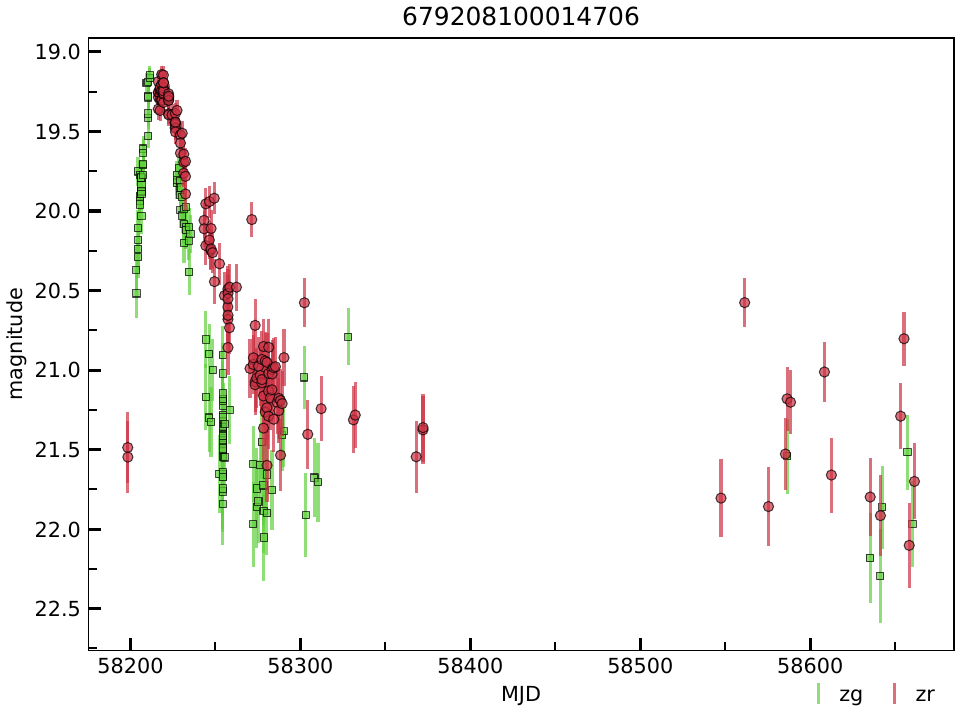}\\b) {\tt SNAD150}: \texttt{679108100003227}~($zg$), \texttt{679208100014706}~($zr$)\\
    \end{minipage}  
    \vfill
    \begin{minipage}{0.49\linewidth}
        \centering
        \includegraphics[scale=0.4]{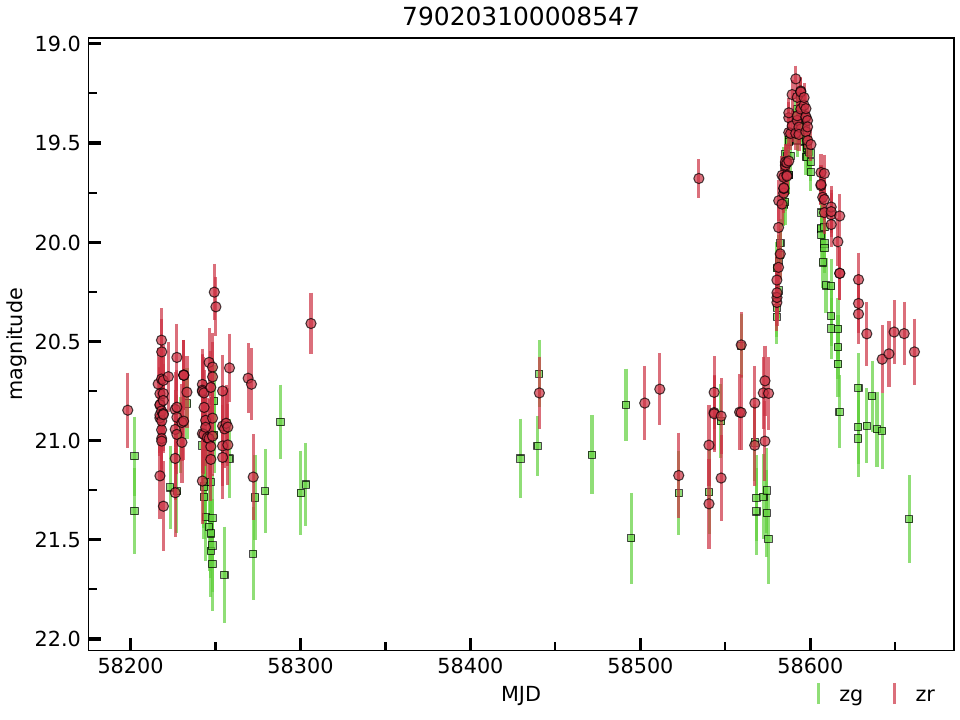}\\c) {\tt SNAD151}: \texttt{790103100000915}~($zg$), \texttt{790203100008547}~($zr$), \texttt{1792109200005099}~($zg$)\\
    \end{minipage}
    \hfill
    \begin{minipage}{0.49\linewidth}
        \centering
        \includegraphics[scale=0.4]{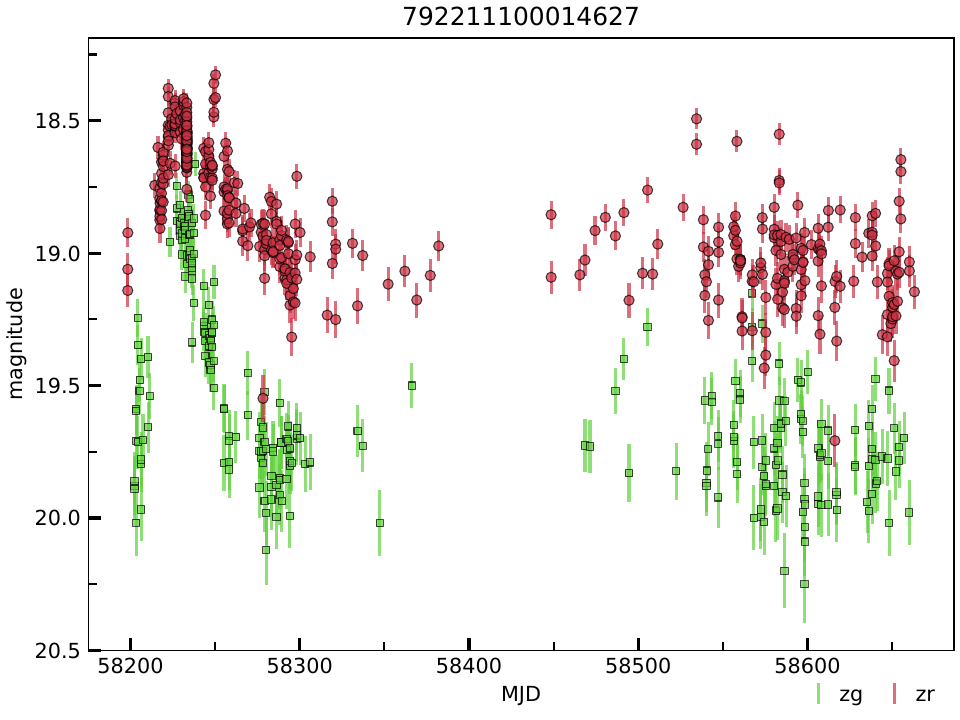}\\d) {\tt SNAD152}: \texttt{792111100012457}~($zg$), \texttt{792211100014627}~($zr$)\\
    \end{minipage}
    \vfill
    \begin{minipage}{0.49\linewidth}
        \centering
        \includegraphics[scale=0.4]{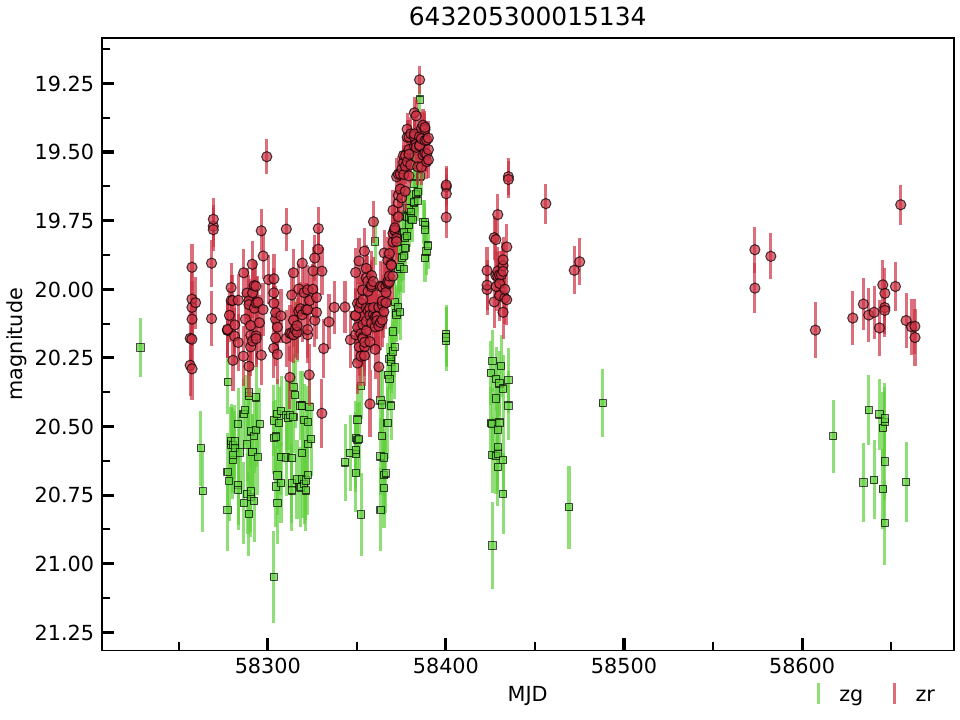}\\e) {\tt SNAD153}: \texttt{643105300009229}~($zg$), \texttt{643205300015134}~($zr$)\\
    \end{minipage}
        \hfill
    \begin{minipage}{0.49\linewidth}
        \centering
        \includegraphics[scale=0.4]{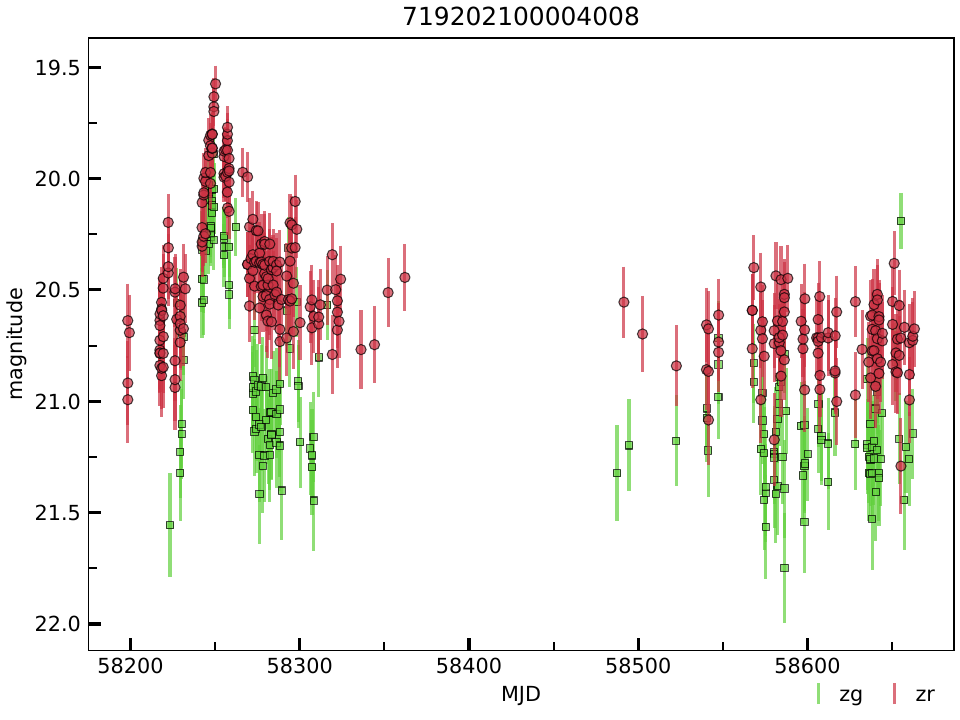}\\f) {\tt SNAD154}: \texttt{719102100006086}~($zg$), \texttt{719202100004008}~($zr$)\\
    \end{minipage}
\end{figure*}
\begin{figure*}
    \begin{minipage}{0.49\linewidth}
        \centering
        \includegraphics[scale=0.4]{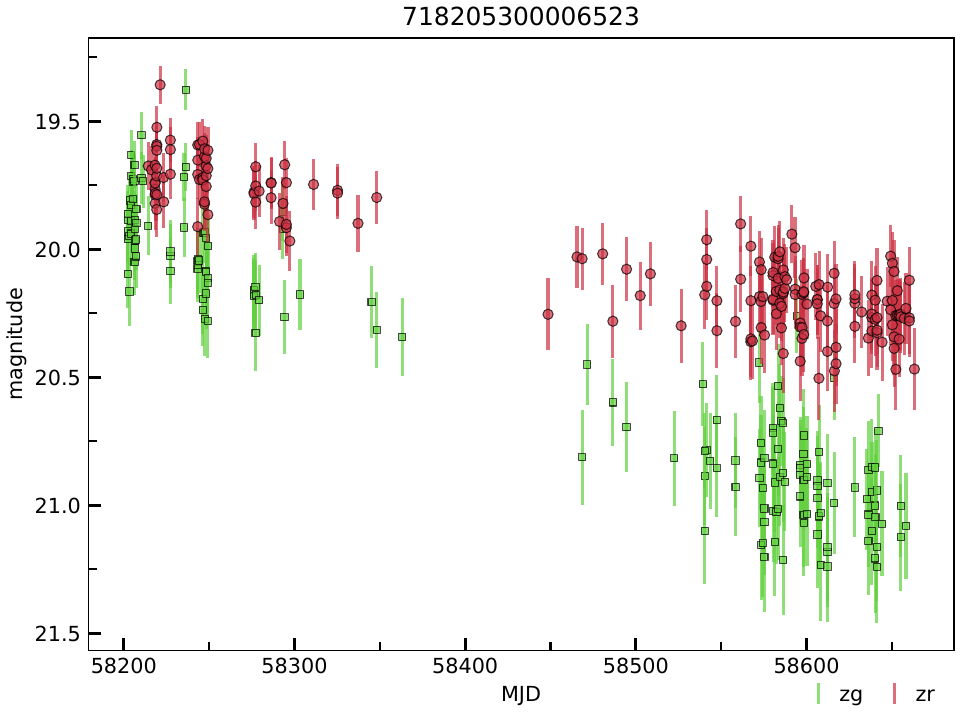}\\g) {\tt SNAD155}: \texttt{718105300007353}~($zr$), \texttt{718205300006523}~($zg$), \texttt{1717111100000191}~($zr$)\\
    \end{minipage}
    \hfill
    \begin{minipage}{0.49\linewidth}
        \centering
        \includegraphics[scale=0.4]{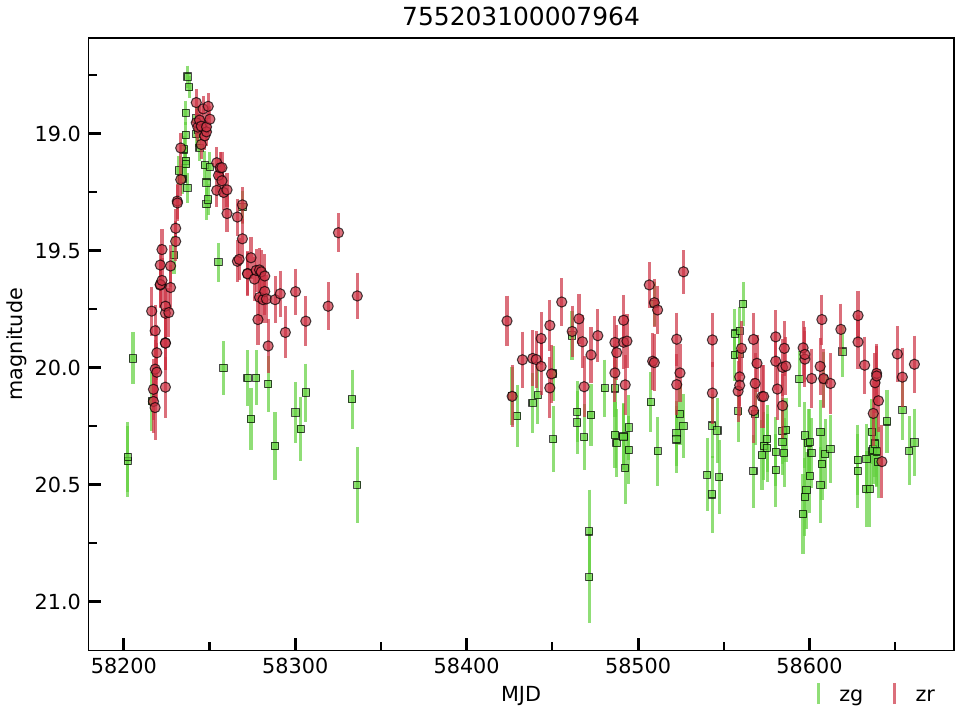}\\h) {\tt SNAD156}: \texttt{755103100026105}~($zg$), \texttt{755203100007964}~($zr$)\\
    \end{minipage}  
    \vfill
    \begin{minipage}{0.49\linewidth}
        \centering
        \includegraphics[scale=0.4]{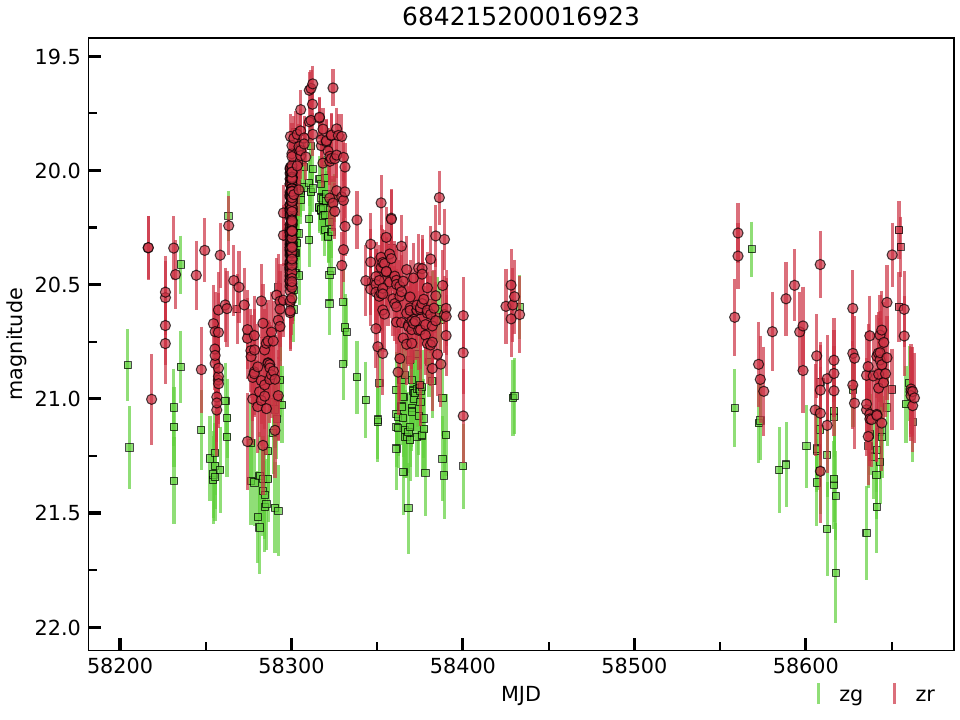}\\i) {\tt SNAD157}: \texttt{684115200018744}~($zg$), \texttt{684215200016923}~($zr$), \texttt{1725206200027715}~($zr$)\\
    \end{minipage}
    \hfill
    \begin{minipage}{0.49\linewidth}
        \centering
        \includegraphics[scale=0.4]{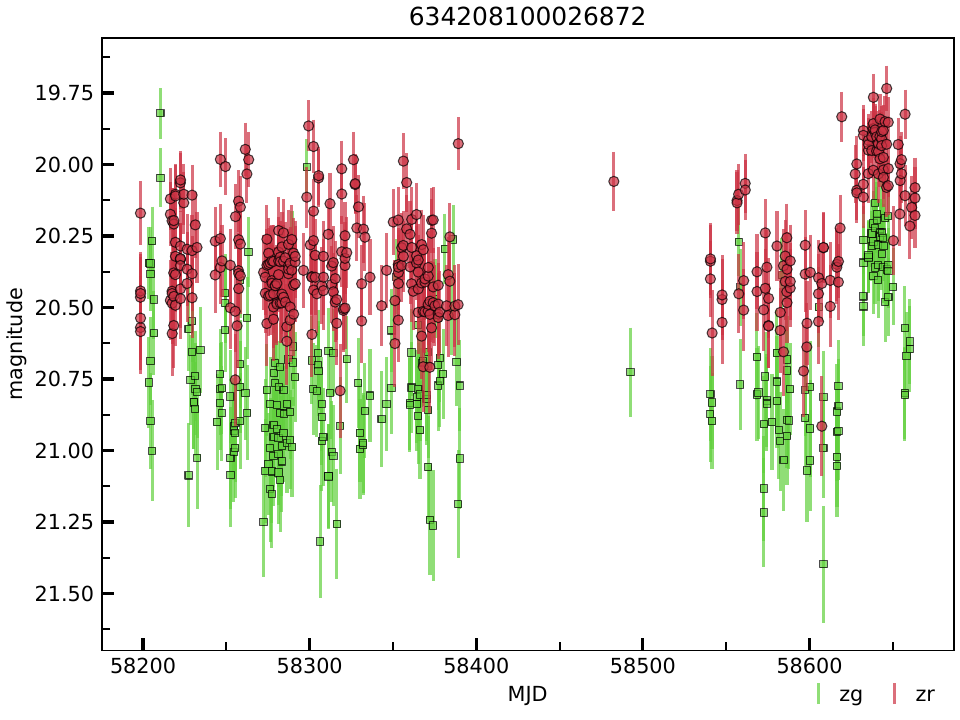}\\j) {\tt SNAD158}: \texttt{634108100006647}~($zg$), \texttt{634208100026872}~($zr$)\\
    \end{minipage}
    \vfill
    \begin{minipage}{0.49\linewidth}
        \centering
        \includegraphics[scale=0.4]{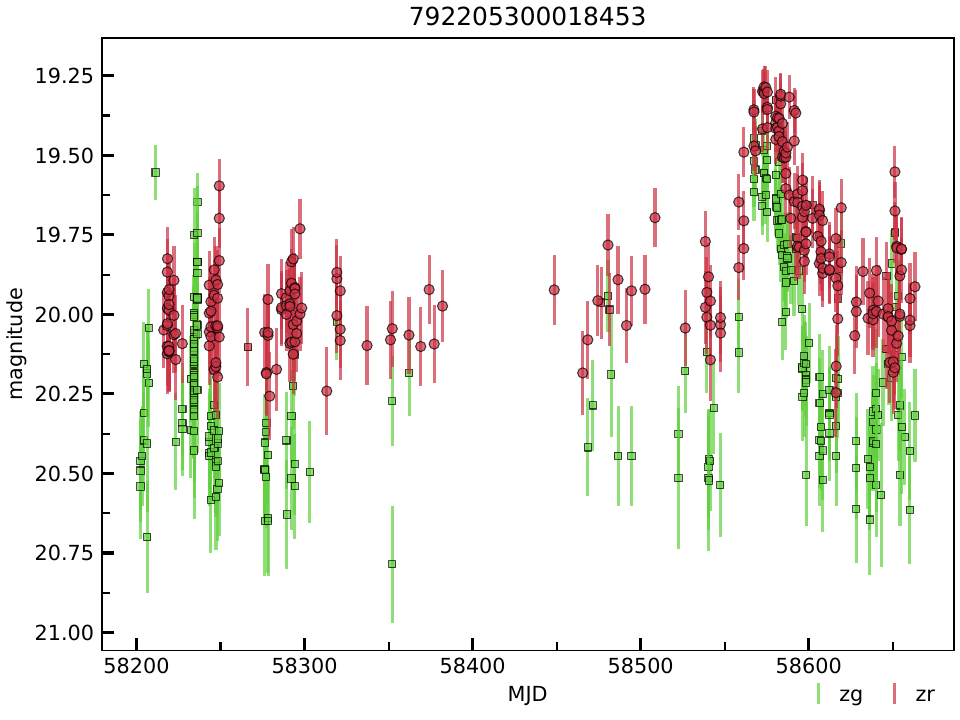}\\k) {\tt SNAD159}: \texttt{792105300007593}~($zg$), \texttt{792205300018453}~($zr$), \texttt{1795111100013935}~($zg$), \texttt{1795211100000534}~($zr$)\\
    \end{minipage}
    \caption{SN/AGN candidate light curves in $zr$- and $zg$-bands within the first 420 days of ZTF DR4, generated with the ZTF \texttt{SNAD} viewer.}
        \label{fig:snad_LC2}
\end{figure*}

\section{Light-curve features}\label{sec:features}
We extract 41 light-curve features for each of $zg-$ and $zr-$passbands using the \texttt{light-curve-feature} Rust crate \citep{Malanchev2021}\footnote{See detailed feature description and package documentation on \url{https://docs.rs/light-curve-feature/0.2.2/}}.
We use the following feature extractors:
\begin{itemize}
    \item a magnitude amplitude;
    \item a magnitude standard deviation;
    \item kurtosis of magnitude and flux distributions, two features;
    \item skews of magnitude and flux distributions, two features;
    \item Anderson--Darling test of normality statistics values for magnitudes and fluxes, two features;
    \item a fraction of observations beyond one/two standard deviations from mean magnitude \citep{DIsanto_etal2016}, two features;
    \item a range of cumulative sums of magnitudes and fluxes \citep{Kim_etal2014}, two features;
    \item a magnitude inter-percentile ranges $2\%-98\%$, $10\%-90\%$ and $25\%-75\%$, three features;
    \item a slope and its error of a linear fit of a magnitude light curve with and without respect to the observation errors, four features;
    \item a ratio of magnitude inter-percentile ranges: 1) $40\%-60\%$ to $5\%-95\%$ and 2) $20\%-80\%$ to $5\%-95\%$ \citep{DIsanto_etal2016}, two features;
    \item a mean magnitude value with and without respect to the observation errors, two features;
    \item a median of the absolute value of the difference between magnitude and median of magnitude distribution \citep{DIsanto_etal2016};
    \item a fraction of observations being within 0.1/0.2 magnitude amplitude from median magnitude \citep{DIsanto_etal2016}, two features;
    \item the maximum deviation of magnitude from median magnitude \citep{DIsanto_etal2016};
    \item a ratio of $5\%-95\%$/$10\%-90\%$ magnitude inter-percentile range to the median magnitude \citep{DIsanto_etal2016}, two features;
    \item signal-to-noise ratio of five largest Lomb--Scargle periodogram peaks, periodogram power standard deviation, and a fraction of periodogram points beyond two/three standard deviations from the mean power value, \citep{Lomb1976,Scargle1982,DIsanto_etal2016}, eight features;
    \item the Stetson $K$ coefficient for magnitudes and fluxes \citep{Stetson1996}, two features;
    \item an excess variance of fluxes \citep{Sanchez2017},
    \item a ratio of flux standard deviation to its mean.
\end{itemize}





\end{document}